\documentclass[draft]{agujournal2019}
\usepackage{url} %this package should fix any errors with URLs in refs.
\usepackage[inline]{trackchanges} %for better track changes. finalnew option will compile document with changes incorporated.
\usepackage{soul}
\usepackage{amsmath}

\usepackage{xcolor}

\DeclareFontFamily{OMS}{oasy}{\skewchar\font48 }
\DeclareFontShape{OMS}{oasy}{m}{n}{%
         <-5.5> oasy5     <5.5-6.5> oasy6
      <6.5-7.5> oasy7     <7.5-8.5> oasy8
      <8.5-9.5> oasy9     <9.5->  oasy10
      }{}
\DeclareFontShape{OMS}{oasy}{b}{n}{%
       <-6> oabsy5
      <6-8> oabsy7
      <8->  oabsy10
      }{}
\DeclareSymbolFont{oasy}{OMS}{oasy}{m}{n}
\SetSymbolFont{oasy}{bold}{OMS}{oasy}{b}{n}

\DeclareMathSymbol{\smallleftarrow}     {\mathrel}{oasy}{"20}
\DeclareMathSymbol{\smallrightarrow}    {\mathrel}{oasy}{"21}
\DeclareMathSymbol{\smallleftrightarrow}{\mathrel}{oasy}{"24}

\usepackage{comment}

\usepackage{float}
\draftfalse

\journalname{JGR: Space Physics}

\begin{document}

\title{Identifying the Growth Phase of Magnetic Reconnection using Pressure-Strain Interaction}

\authors{M.~Hasan Barbhuiya\affil{1}, Paul A.~Cassak\affil{1}, Alex Chasapis\affil{2}, Michael A.~Shay\affil{3}, Giulia Cozzani\affil{4}, Alessandro Retin\`o \affil{5}}

\affiliation{1}{Department of Physics and Astronomy and the Center for KINETIC Plasma Physics, 
West Virginia University, Morgantown, WV 26506, USA
}

\affiliation{2}{ Laboratory for Atmospheric and Space Physics, University of Colorado Boulder, Boulder, CO 80303, USA}

\affiliation{3}{Bartol Research Institute, Department of Physics and Astronomy, University of Delaware, Newark, Delaware, 19716, USA}

\affiliation{4}{Department of Physics, University of Helsinki, Helsinki, Finland}

\affiliation{5}{Laboratoire de Physique des Plasmas (LPP), UMR7648, CNRS, Sorbonne Universit{\'e}, Universit{\'e} Paris-Saclay, Observatoire de Paris, Ecole Polytechnique Institut Polytechnique de Paris, Paris, France}

\correspondingauthor{M. Hasan Barbhuiya}{mhb0004@mix.wvu.edu}

%% Keypoints, final entry on title page.

%  List up to three key points (at least one is required)
%  Key Points summarize the main points and conclusions of the article
%  Each must be 100 characters or less with no special characters or punctuation and must be complete sentences

\begin{keypoints}
\item Separatrices open during the reconnection growth phase causing a negative pressure-strain interaction signal.

\item The negative features are at the flanks of a strong positive signal caused by dipolarization front.

%\item \textcolor{orange}{This signal created by diverging electron flows identifies the growth phase of reconnection.} 

%\item Simulations confirm it as a marker of the growth phase and is strong enough to be detected by NASA's MMS mission.
%Simulations confirm this marker and that it is strong enough to be detected by NASA's MMS mission. 
\item Simulations confirm negative features are markers of the growth phase and are detectable by NASA MMS mission.
\end{keypoints}

\begin{abstract}
Magnetic reconnection often initiates abruptly and then rapidly progresses to a nonlinear quasi-steady state. While satellites frequently detect reconnection events, ascertaining whether the system has achieved steady-state or is still evolving in time remains challenging. Here, we propose that the relatively rapid opening of reconnection separatrices within the electron diffusion region (EDR) serves as an indicator of the growth phase of reconnection. The opening of the separatrices is produced by electron flows diverging away from the neutral line downstream of the X-line and flowing around a dipolarization front.  This flow pattern leads to characteristic spatial structures in pressure-strain interaction that could be a useful indicator for the growth phase of a reconnection event. We employ two-dimensional particle-in-cell numerical simulations of anti-parallel magnetic reconnection to validate this prediction. We find that the signature discussed here, alongside traditional reconnection indicators, can serve as a marker of the growth phase. This signature is potentially accessible using multi-spacecraft single-point measurements, such as with NASA’s Magnetospheric Multiscale (MMS) satellites in Earth’s magnetotail. Applications to other settings where reconnection occurs are also discussed.
\end{abstract}

\section*{Plain Language Summary}
Magnetic reconnection is a process where magnetic energy is rapidly converted into energy in the surrounding charged particles. It plays an important role in geomagnetic substorms, which cause aurora and the harmful side-effects of space weather. It is difficult to tell using satellite missions when this process is just starting or has been going for a while. In this study, we suggest that a key feature of reconnection can help identify when the process is in its early, growing stage.  As reconnection starts, a jet of electrons moves away from the reconnection site and gets diverted, causing the magnetic fields participating in reconnection to open wider. This expansion of the electron jet is measurable using a quantity called the pressure-strain interaction. Using supercomputer simulations, we confirm that this quantity, along with other common signs of reconnection, can be used to infer if reconnection has just started or not and should be strong enough to be measured by satellites -- especially NASA's Magnetospheric Multiscale (MMS) mission. The findings could help scientists better understand reconnection events using spacecraft data and apply this understanding to other space environments where reconnection occurs.

\section{Introduction}
\label{sec:intro}

Magnetic reconnection is a process that converts magnetic energy to bulk flow energy and internal energy of the plasma \cite{Zweibel09}. It occurs in near-Earth plasma systems such as the dayside magnetopause \cite{Paschmann79,Burch_and_Phan_GRL_2016} and the magnetotail \cite{McPherron79,Oieroset01,Angelopoulos08}. Given its crucial role in influencing magnetospheric dynamics and its implications for space weather \cite{Pulkkinen07}, several satellite missions, including Cluster \cite{CLUSTER_mission_2001}, Time History of Events and Macroscale Interactions during Substorms (THEMIS) \cite{Angelopoulos2009}, and Magnetospheric Multiscale (MMS) \cite{Burch16}, have focused on reconnection as a target of its in-situ observations and analysis. 

Reconnection in Earth's magnetotail characteristically evolves in time marked by reasonably distinct evolutionary phases ({\it e.g.,} \cite{Bhattacharjee04}).  At onset, reconnection begins, often rather slowly on ion cyclotron time scales, perhaps through a tearing instability [{\it e.g.}, \cite{sitnov2010tearing,Liu_2014_JGR}]. Then, the reconnection rate explosively increases, during what has been called the growth phase \cite{Bhattacharjee04}.  Assuming steady upstream conditions, the reconnection rate then saturates, leading to a steady-state phase [{\it e.g.}, \cite{priest1986new,shay1998role,Birn_2005_GRL}] that
lasts at least tens of inverse ion cyclotron times and continues until the available magnetic energy is exhausted and reconnection stops. During the steady-state phase, the reconnection rate fluctuates around a specific value and thus, on average, does not change significantly. At least in numerical simulations, the steady-state phase is often preceded by what we refer to as the overshoot phase, where the reconnection rate at the end of the growth phase is higher than the steady-state value and then relaxes down to it \cite{shay2007two,Payne_JGR_2024}.  Reconnection in the solar corona associated with flares and coronal mass ejections is expected to evolve through a similar set of phases as in Earth's magnetotail \cite{priest2002magnetic,dahlburg2005explanation,reeves2008posteruptive}. 

There are many signatures used to identify the occurrence of reconnection in satellite observations, including ion \cite{Paschmann79} and electron \cite{Phan07,Torbert18} jets, as well as the presence of Hall magnetic fields and electric fields \cite{Nagai01,Oieroset01,Mozer02,Wygant2005_JGR} in a region where a component of the magnetic field reverses, {\it i.e.}, current sheets. The unprecedented measurement resolution provided by MMS has facilitated routine observations of non-Maxwellian velocity distribution functions during reconnection that have also become key diagnostics of the process, such as those observed in Earth's magnetopause \cite{Burch16b} and magnetotail \cite{Torbert18,Xinmin_li_2019_GRL}. 
While identifying the phases of reconnection in two-dimensional (2D) numerical simulations is straightforward using flux function techniques [{\it e.g.}, \cite{shay1998role}], it is extremely challenging to ascertain what phase of reconnection occurs when observed in-situ with satellites during current sheet crossings.

In the MMS era, research on magnetotail reconnection onset has largely focused on electron dynamics \cite{Hesse_2001_JGR,Liu_2014_JGR,Spinnangr_2022_GRL}. A recent study suggested the possible detection of reconnection onset \cite{Genestreti_2023_JGR_Onset1,Genestreti_2023_JGR_Onset2}, marked by rapid thinning of the cross-tail current sheet below the ion inertial scale \cite{Nakamura_2006_SSR,Sitnov_2019_SSR} and the subsequent onset of the electron tearing mode \cite{Liu_2014_JGR}. Another observational study \cite{Hubbert_electron-only_2022_JGR}, along with simulations \cite{Lu_electron-only_2022_JGR}, examined current and plasma sheet properties during reconnection onset. To identify the growth phase in simulations, a recent study proposed tracking the evolution of the non-ideal reconnection electric field and the divergence of the Poynting flux \cite{Payne_JGR_2024}.

In the present study, we identify a different physical marker of the reconnection growth phase that can be applied to satellite observations. This marker is associated with the opening out of the magnetic field separatrix in the downstream portion of the electron diffusion region (EDR) during the growth phase. The opening out of the reconnected magnetic field is imperceptible during the onset phase but occurs relatively rapidly during the growth phase, then is stabilized at the beginning of the steady-state phase. When measured in tandem with other reconnection metrics, we expect it to be a signature of the growth phase. Directly identifying the opening out of the reconnected magnetic field observationally would be difficult, but it is associated with a diverging electron flow downstream of the X-line within the EDR. This divergence is identifiable using the pressure-strain interaction \cite{del_sarto_pressure_2016,del_sarto_pressure_2018,Yang17,yang_PRE_2017}, a scalar quantity that describes the change in internal energies as a result of converging/diverging flow and/or sheared flow \cite{Cassak_PiD1_2022,Cassak_PiD2_2022}. The quantity has been measured using MMS \cite{Chasapis18,Bandyopadhyay20}. Therefore, this identifier has desirable properties as a measure of the growth phase, namely uniqueness ({\it i.e.,} a signature absent in other phases), measurability with satellites, and is reference frame and coordinate system independent (\textit{i.e.,} a scalar, rather than a vector that can be calculated without identifying a coordinate system associated with reconnection, such as boundary normal (LMN) coordinates, which introduces uncertainty \cite{genestreti2018accurately}).  

In a previous study \cite{Barbhuiya_PiD3_2022}, it was hypothesized that the pressure-strain interaction signature could serve as a marker for the growth phase of reconnection. This hypothesis was based on a comparison of their particle-in-cell (PIC) simulations in the growth phase with simulations in the steady-state phase \cite{Pezzi21}. However, the simulation domain used in Ref.~\cite{Barbhuiya_PiD3_2022} was too small to achieve a steady state. In the present study, we establish with larger simulations that the proposed signature is a viable identifier for the growth phase.

The layout of this paper is as follows.  The theory is discussed in Sec.~\ref{sec:theory}. Section~\ref{sec:simulation} describes the numerical simulations carried out in this study.  The simulation results are discussed in Sec.~\ref{sec:results}. Section~\ref{sec:conclusion} summarizes the results and discusses further implications.

\begin{figure}
    \begin{center}
    \includegraphics[width=4.5in]{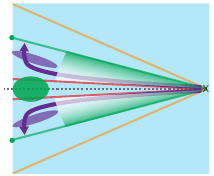}
    \caption{A sketch of the time evolution of the magnetic fields inside the electron diffusion region during the onset and growth phases of magnetic reconnection. The left half of the EDR is sketched, with the X-line marked by the ``x'' on the right hand side. The gold lines denote the separatrices when reconnection is in a steady-state, and the blue shading is the EDR in the steady-state.  The red lines denote the separatrices in the onset phase, when they open out slowly. The green shading leading to the green line depicts the separatrices opening out more rapidly in the growth phase.  During this phase, the electron outflow jets (the purple arrows) deflect outward in the current sheet normal direction. This produces diverging electron bulk flow, giving rise to a negative electron pressure-strain interaction (the purple ellipses). Sandwiched in between is a strong positive electron pressure-strain interaction (the green ellipse), associated with compression in a dipolarization front.  
    The distance between the ``x'' and the green circles is defined as $L_{out,\parallel}$ as described in the text.  The negative pressure-strain interaction signature disappears in the steady-state phase when the separatrix angle reaches its maximum represented by the gold line.}
    \label{fig:blueellipsesevol}
    \end{center}
\end{figure}

\section{Theory}
\label{sec:theory}

As discussed in the previous section, the opening of the separatrix as reconnection progresses is the key physics behind the proposed physical marker for the growth phase of reconnection.  We first describe the evolution qualitatively, using the sketch in Fig.~\ref{fig:blueellipsesevol} which contains the left half of the EDR, with the X-line at the center of its right edge denoted by ``x.'' The gold lines denote the separatrix when the reconnection is in the steady-state. The EDR when the reconnection is in the steady-state is denoted as the shaded light blue rectangle.  During the onset phase, the separatrices, sketched as the red lines, start opening out and they do so relatively slowly. In the growth phase, the separatrix opens relatively rapidly in time. This rapid opening is depicted as the green shading leading to the green line in the figure. This is caused by the deflection of the electrons in the exhaust jet (denoted by the purple arrows) around the pre-existing current sheet and the newly reconnected magnetic field lines near the outflow edges of the EDR. Therefore, the deflection of the electron flow is likely to occur alongside dipolarization fronts in Earth's magnetotail \cite{Ohtani_2004_JGR, runov2009themis,sitnov2009dipolarization,fu2020magnetotail}; the diverging flows initially manifests slightly downstream of the dipolarization front and are displaced slightly outward in the current sheet normal direction.  During the steady-state, the separatrices settle at their steady-state locations (the gold lines). Note, the separatrices outside the EDR continue to evolve in time after the separatrices in the EDR have reached their steady-state positions.
  
The key aspect for a physical marker to determine that the reconnection is in the growth phase is the electron flows diverging away from the neutral line in the region surrounding the downstream edge of the EDR.  Diverging electron flow implies the electrons are undergoing expansion. The diverging electron flow is identifiable as a negative region in a plot of the electron pressure-strain interaction, as was discussed previously \cite{Barbhuiya_PiD3_2022}.

To see this analytically, electron pressure-strain interaction is defined as $-{\bf P}_e:{\bf S}_e = -({\bf P}_e \cdot \nabla) \cdot {\bf u}_e$, where ${\bf P}_e$ is the electron pressure tensor with elements $P_{e,jk} = m_e \int v_{e j}^\prime v_{e k}^\prime f_e d^3v$, $k_B$ is Boltzmann's constant, $f_e$ is the electron phase space density, ${\bf v}_e^\prime = {\bf v} - {\bf u}_e$ is the random (peculiar) velocity, ${\bf v}$ is the velocity-space coordinate, $n_e = \int f_e d^3v$ is the electron number density, ${\bf u}_e = (1/n_e) \int f_e {\bf v} d^3v$ is the electron bulk flow velocity, $m_e$ is the electron mass, and ${\bf S}_e$ is the symmetric part of the strain rate tensor $ \nabla {\bf u}_e$ with elements $S_{e,jk}=(1/2)(\partial u_{ek}/\partial r_j + \partial u_{ej}/\partial r_k)$. Electron pressure-strain interaction contains both, \textit{i.e,} pressure dilatation, $-P_e (\nabla \cdot {\bf u}_e)$, where $P_e = (1/3) {\rm tr}({\bf P}_e)$ and incompressible contributions called ${\rm Pi-D}$ \cite{del_sarto_pressure_2016,del_sarto_pressure_2018,Yang17,yang_PRE_2017}.

The negative pressure-strain interaction signal associated with the diverging electron flow should be localized in the downstream region of the EDR displaced away from the neutral line but inside the region bounded by the separatrices. Its extent along the current sheet normal direction should be no larger than an electron inertial length, and its extent in the outflow direction should be at most a few electron inertial lengths \cite{Shay_1999_GRL}.  When occurring alongside a dipolarization front, it is highly likely that the region between the two negative electron pressure-strain interaction signals exhibits a relatively strong positive electron pressure-strain interaction signal, as dipolarization fronts are associated with strong compression \cite{runov2011dipolarization,BarbhuiyaRing22}. This is depicted as the green ellipse near the left-center of the figure. We argue that the negative signals without a strongly positive signal near their center would be ambiguous, so measuring a strongly positive part is important to claim an event is in the growth phase.

The negative pressure-strain features within the EDR vanish when the separatrix reaches its steady-state opening angle. However, in the ion diffusion region (IDR) outside the EDR, the separatrices may still be adjusting to their steady-state positions. We predict that negative electron pressure-strain features will emerge further downstream in the IDR as it evolves. As the separatrix opening angle in the IDR reaches its asymptotic value further downstream, the negative pressure-strain signal will appear to move downstream.

\section{Numerical Simulations} 
\label{sec:simulation}

We use the massively parallel PIC code {\tt p3d} \cite{zeiler:2002} for a 3D velocity-space and 2.5D position-space magnetic reconnection simulation. The simulation employs a relativistic Boris particle stepper for macro-particle evolution \cite{birdsall91a} and the trapezoidal leapfrog method for electromagnetic fields \cite{guzdar93a}. Poisson's equation is enforced using the multigrid method \cite{Trottenberg00}. Periodic boundary conditions are used in both spatial directions.

Simulation data are presented in normalized units. Lengths are normalized to the ion inertial scale $d_{i0} = c/\omega_{pi0}$, where $c$ is the speed of light, and $\omega_{pi0} = (4 \pi n_0 q_i^2 /m_i)^{1/2}$ is the ion plasma frequency with reference density $n_0$, ion charge $q_i$, and ion mass $m_i$. Magnetic fields are normalized to $B_0$. Velocities are normalized to the Alfv\'en speed $c_{A0} = B_0/(4 \pi m_i n_0)^{1/2}$. Electric fields are normalized to $E_0=B_0 c_{A0}/c$. Times are normalized to $\Omega_{ci0}^{-1} = (q_i B_{0} / m_{i} c)^{-1}$. Temperatures are normalized to $m_i c_{A0}^2/k_B$, and power densities are in units of $(B_0^2/4 \pi) \Omega_{ci0}$. 

For numerical efficiency, we use non-physical values for the speed of light ($c=15$) and the electron-to-ion mass ratio ($m_e/m_i = 0.04$). These choices do not affect the relevant physics. The smallest length scale is the electron Debye length, and we use a grid length $\Delta=0.0125$. The smallest time scale is the inverse electron plasma frequency and we employ a particle time step $\Delta t=0.001$ and a field time step of $0.0005$. The electric field is cleaned every 10 particle time steps. The simulation domain is $L_x \times L_y = 102.4 \times 51.2$, which is sufficient to ensure both growth and steady-state phases take place and there are 200 weighted particles per grid (PPG) initially.

We next discuss the initialization of the simulation. We employ a double $\tanh$ magnetic field profile, 
\[
    B_x(y) = \tanh{\left(\frac{y-L_y/4}{w_0}\right)} - \tanh{\left(\frac{y-3L_y/4}{w_0}\right)} -1,
\]
and zero guide magnetic field $B_z$, where $w_0=0.5$ is the half-thickness of the current sheet. Electron and ion density profiles are 
\[
    n(y) = \frac{1}{2(T_{e}+T_{i})} \left[ {\rm sech}^{2}\left(\frac{y-L_y/4}{w_0}\right) + {\rm sech}^{2}\left(\frac{y-3L_y/4}{w_0}\right) \right] + n_{up},
\]
where the initial asymptotic upstream plasma density $n_{up}=0.2$, and the difference between the peak current sheet number density and $n_{up}$ is the reference density $n_0$. The electron and ion temperatures are $T_{e}=1/12$ and $T_{i}=5T_{e}$, respectively, with electron and ion out-of-plane drift speeds satisfying $u_{e,z}/T_e = -u_{i,z}/T_i$. Both species are loaded as single drifting Maxwellian velocity distribution functions instead of two populations for the Harris sheet kinetic equilibrium \cite{Harris62}. Reconnection is initiated by seeding X- and O-line pairs with a magnetic field perturbation given by
\[
\delta B_x = -B_{pert} \sin \left(\frac{2 \pi x}{L_x} \right) \sin\left(\frac{4\pi y}{L_y} \right)
\]
and 
\[
\delta B_y =  B_{pert} \left(\frac{L_y}{2 L_x}\right) \cos\left(\frac{2 \pi x}{L_x}\right) \left[1-\cos\left(\frac{4\pi y}{L_y}\right)\right],
\]
where $B_{pert} = 0.05$.  

In this study, we use a lower PPG than in a previous investigation \cite{Barbhuiya_PiD3_2022}, which used identical plasma parameters with a higher PPG (25,600) in a smaller domain. To counter the increased noise in the 2D plots, we recursively smooth the raw simulation data which involves twelve iterations of smoothing over four cells, calculating spatial derivatives, and another twelve iterations of smoothing over four cells. We selected these smoothing parameters through systematic testing to ensure the 2D power density plots during the reconnection growth phase resemble those in the previous study.

\section{Results} 
\label{sec:results}

\begin{figure}
\begin{center}
    \includegraphics[width=3.2in]{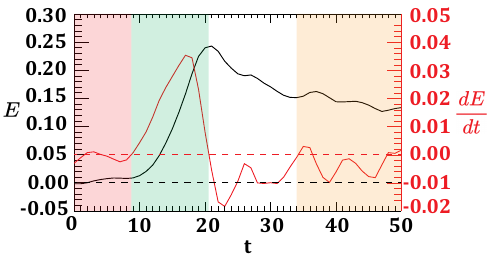}
    \caption{Time evolution of the reconnection rate $E$ (black) and its time derivative $dE/dt$ (red) as a function of time $t$ for the lower current sheet. The dashed lines mark where the associated quantity is zero. The red, green, white, and gold-shaded regions approximately demarcate the onset, growth, overshoot, and steady-state phases, respectively.}
    \label{fig:rratevstime}  
\end{center}
\end{figure}

We start with an overview of the simulation, where we link the time evolution of the reconnection rate $E$ with the phases of reconnection, where $E$ is calculated using a flux function technique, with $E$ being the time derivative of the flux function $\psi$, where the in-plane magnetic field is given by ${\bf \hat{z}} \times \nabla \psi$. Figure~\ref{fig:rratevstime} shows the reconnection rate $E$ in black and its time derivative $dE/dt$ in red using the vertical axes on the left and right side, respectively as a function of time $t$ for the lower current sheet centered at $y=L_y/4=12.8$. The time period shaded in red up to $t \simeq 9$ approximately denotes the onset phase when $E$ is small and $dE/dt$ is close to zero. 
Next in green shading is the growth phase up to time $t \simeq 20$ when $E$ most rapidly increases in time and $dE/dt>0$. We refer to the time at which $dE/dt$ is at its maximum, \textit{i.e.,} $t=18$, as the ``peak growth time.'' 
The growth phase is followed by the overshoot phase shaded in white that begins right after the peak in $E$ and when $dE/dt$ reaches zero and goes until $t \simeq 34$. 
Finally, the steady-state phase is shaded in gold when $E$ is relatively constant and $dE/dt$ is small for approximately 15 $\Omega^{-1}_{ci0}$. These results are qualitatively consistent with previous work on the nonlinear time evolution of reconnection [{\it e.g.}, \cite{Bhattacharjee04}].

\begin{figure}
\begin{center}
    \includegraphics[width=2.75in]{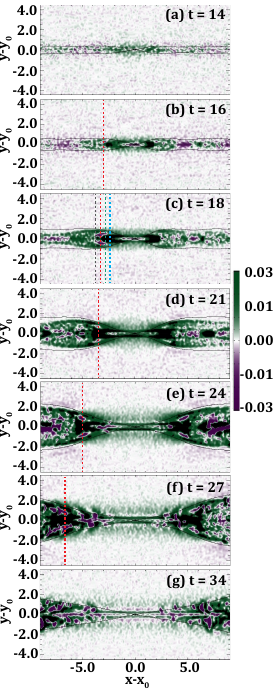}
    \caption{2D spatial structure of $-{\bf P}_e:{\bf S}_e$ at (a)-(g) $t=14,~16,~18,~21,~24,~27$, and 34, respectively. The color bar is saturated to make weaker values more visible; the minimum and maximum values for the different times are (a) ($-0.029,0.022$), (b) ($-0.089,0.029$), (c) ($-0.12,0.027$), (d) ($-0.28,0.31$), (e) ($-0.21,0.29$), (f) ($-0.19,0.20$), and (g) ($-0.22,0.37$). Black solid lines denote the separatrices at each time. Vertical red dashed lines through the approximate center of the $-{\bf P}_e:{\bf S}_e <0$ features are in panels (b) to (f) (demonstrating its motion away from the X-line) along with gray, green and blue dashed lines at the peak growth phase time in panel (c) along which cuts are shown in Fig.~\ref{fig:1Dcut_t18}.} 
    \label{fig:2D_PS_NegEllipses}
\end{center}
\end{figure}

To assess the plausibility of using the pressure-strain interaction to identify the growth phase, we show 2D plots of the spatial variation of the electron pressure-strain interaction as a function of time in Fig.~\ref{fig:2D_PS_NegEllipses}. We use a saturated color bar to more clearly display the structure of the electron pressure-strain interaction.  At each time, we shift the coordinate system to be relative to the X-line location $(x_0,y_0)$. Separatrices are in solid black. Panels (a)-(c) show data from $t=14, 16$, and $18$, respectively, during the growth phase, followed by panels (d)-(f) from $t=21, 24$, and $27$, respectively, during the overshoot phase, followed by panel (g) that shows data at $t=34$ during the steady-state phase. 

Early in the growth phase at $t=14$, we observe a weak positive signal near the EDR in Fig.~\ref{fig:2D_PS_NegEllipses}(a), but coherent negative features are not observed. As seen in Fig.~\ref{fig:2D_PS_NegEllipses}(b), by $t \simeq 16$, two coherent negative pressure-strain interaction features (in purple) emerge centered around $(x-x_0,y-y_0) \simeq (-3.0,\pm 0.25)$ which is marked by a vertical red dashed line. This negative electron pressure-strain interaction signal is located near a strong positive signal (due to compression at the dipolarization front) centered at $(x-x_0,y-y_0) \simeq (-2.5,0)$.

Looking at the peak growth time of $t=18$ in Fig.~\ref{fig:2D_PS_NegEllipses}(c), two regions of negative electron pressure-strain interaction are apparent in the region near $(x-x_0,y-y_0)\approx (-3.3,\pm 0.5)$, also marked by a vertical red line. For this simulation, the electron inertial scale based on the upstream density $n_{up}$ is $d_e \simeq 0.45 \ d_{i0}$. The approximate centers of these $-{\bf P}_e:{\bf S}_e$ features are displaced approximately $1 \ d_e$ from the neutral line.  In addition, since the expected extent in the outflow direction of the EDR is $5-10~d_e \simeq 2.2-4.4 \ d_{i0},$ the features are near the downstream edge of the EDR, as expected from the discussion in Sec.~\ref{sec:theory}.  The structure looks similar to that seen in Ref.~\cite{Barbhuiya_PiD3_2022}, which employed a smaller simulation domain (see their Fig.~2). We note that there is no clear counterpart to the $-{\bf P}_e:{\bf S}_e$ feature to the right of the X-line at $t = 16$ and $18$. By inspection of the simulation results, we find there is a secondary island to the right of the X-line that formed during the onset phase prior to $t=14$. We hypothesize that it interfered with a clean signal to the right of the X-line at $t = 16$ and $18$. However, the feature does arise to the right of the X-line at later times as seen in Fig.~\ref{fig:2D_PS_NegEllipses}(d)-(f).

We next discuss the results during the overshoot phase in panels (d)-(f). As in Fig.~\ref{fig:2D_PS_NegEllipses}(c), we draw a vertical red dashed line through the approximate center of the negative $-{\bf P}_e:{\bf S}_e$ features in these panels.  The locations of the vertical lines illustrate that the negative $-{\bf P}_e:{\bf S}_e$ features have an apparent motion downstream, away from the X-line, that happens outside the downstream edge of the EDR, as anticipated in Sec.~\ref{sec:theory}. 

Fig.~\ref{fig:2D_PS_NegEllipses}(g) is at $t=34$ 
once the system reaches the steady-state phase. Here, the coherent negative $-{\bf P}_e:{\bf S}_e$ signal with a strongly positive pressure-strain interaction signal sandwiched in between are no longer visible, as anticipated in Sec.~\ref{sec:theory}. This demonstrates the utility of this signature in determining the phase of reconnection.

\begin{table}
    \caption{Properties of the approximate centers of the negative $-{\bf P}_e:{\bf S}_e$ features as a function of time. The columns give the time $t$, the $x$ and $y$ coordinates of the approximate centers of the negative $-{\bf P}_e:{\bf S}_e$ regions relative to the X-line position $(x_0,y_0)$. Rows 1, 3, 6, 9, and 11 use fonts of the same color scheme as used in the 1D cuts shown in Fig.~\ref{fig:1Dcutsallstages}.}
 \centering
 \begin{tabular}{c c c}
\hline
$t$  & $x-x_0$ & $y-y_0$ \\ \hline
\textcolor[HTML]{88b94f}{16} & \textcolor[HTML]{88b94f}{-3}      & \textcolor[HTML]{88b94f}{-0.25}     \\ 
17 & -3.1    & -0.4      \\ 
\textcolor[HTML]{1b9e77}{18} & \textcolor[HTML]{1b9e77}{-3.3}    & \textcolor[HTML]{1b9e77}{-0.5}       \\ 
19 & -3.3    & -0.6      \\ 
20 & -3.3    & -0.7      \\ 
\textcolor[HTML]{d95f02}{21} & \textcolor[HTML]{d95f02}{-3.6}    & \textcolor[HTML]{d95f02}{-0.8}        \\ 
22 & -4    & -0.8        \\ 
23 & -4.5    & -0.8      \\ 
\textcolor[HTML]{7570b3}{24} & \textcolor[HTML]{7570b3}{-4.5}     & \textcolor[HTML]{7570b3}{-0.9}       \\ 
25 & -5.8    & -1         \\ 
\textcolor[HTML]{e7298a}{27} & \textcolor[HTML]{e7298a}{-6.7}    & \textcolor[HTML]{e7298a}{-1}           \\  \hline
\end{tabular}
\label{tab:NegEllipseData}
\end{table}

\begin{figure}
\begin{center}
    \includegraphics[width=3.2in]{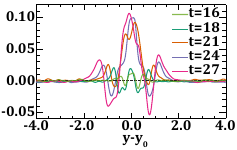}
    \caption{1D cuts of the electron pressure-strain interaction at the approximate centers of the $-{\bf P}_e:{\bf S}_e <0$ signals along the red lines in panels (b)-(f) of Fig.~\ref{fig:2D_PS_NegEllipses}. Light green, dark green, orange, purple and magenta line,s are for $t=16$, $t=18$, $t=21$, $t=24$, $t=27$, respectively. This plot  demonstrates the increase in the amplitude of the $-{\bf P}_e:{\bf S}_e <0$ signal in time}.
    \label{fig:1Dcutsallstages}
\end{center}
\end{figure}

We revisit the apparent motion of the negative electron pressure-strain interaction features. We tabulate the locations ($x-x_0,~y-y_0$) of the approximate centers of the negative $-{\bf P}_e:{\bf S}_e$ features in the second and third columns of Table \ref{tab:NegEllipseData} as a function of time $t$ in the first column from the growth phase to the overshoot phase.  
We find that the average speed of the apparent motion in the outflow direction between $t=16$ and 27 is about $0.34 \ c_{A0} \approx 0.16 \ v_{A} \approx 0.03 \ v_{Ae}$ in the simulations, which could potentially be compared with satellite data.
There is also an apparent motion of the negative features in the current sheet normal direction. This is shown vividly in Fig.~\ref{fig:1Dcutsallstages}, which displays 1D vertical cuts along the red dashed lines in Fig.~\ref{fig:2D_PS_NegEllipses} panels (b)-(f) in light green, dark green, orange, purple, and magenta, respectively. The negative portion of each of the line plots appears to move in time towards higher $|y-y_0|$. 

We quantify the temporal evolution of the amplitude of the negative electron pressure-strain interaction signal in Fig.~\ref{fig:1Dcutsallstages}. It %also reveals that the amplitude of the negative electron pressure-strain interaction 
increases in time, going from $\approx 0.01$ at $t=16$ to $\simeq 0.05$ at $t=27$. 
%During the overshoot phase, b
The 1D cuts demonstrate that the  positive pressure-strain interaction signal associated with the dipolarization front is much stronger than the two negative features. % is a strongly positive pressure-strain interaction signal associated with the dipolarization front that aligns with the  negative features in this phase; this behavior is as anticipated in Sec.~\ref{sec:theory}. 
This strong positive $-{\bf P}_e:{\bf S}_e$ signal is not seen in the 1D cuts taken during the growth phase at $t=16$ and 18 (the light and dark green lines in Fig.~\ref{fig:1Dcutsallstages}) because the positive and negative signals are displaced in the $x$ direction.%, as discussed earlier.

%\textcolor{orange}{We tabulate the locations ($x-x_0,~y-y_0$) of the approximate centers of the negative $-{\bf P}_e:{\bf S}_e$ features in the second and third columns of Table \ref{tab:NegEllipseData} as a function of time $t$ in the first column from the growth phase to the overshoot phase. The last column displays the angle $\phi$.  As discussed in Sec.~\ref{sec:theory}, the negative features of $-{\bf P}_e:{\bf S}_e$ convect downstream with time, while moving away from the neutral line during the growth phase while $\phi$ rapidly increases.  After $t = 21$ as the overshoot phase begins, the $-{\bf P}_e:{\bf S}_e<0$ features continue to move downstream, and their motion away from the neutral line starts to slow down as they move outside the EDR; this manifests as $\phi$ first plateauing and then decreasing. After $t = 27$, the negative features of $-{\bf P}_e:{\bf S}_e$ convect further downstream (not shown) and disappear. The speed of the apparent motion in the downstream direction during the overshoot phase is about 0.16 $v_{A}$ in the simulations.}

\section{Discussion and Conclusion} 
\label{sec:conclusion}

%\textcolor{orange}{We present a physical marker, {\it i.e.,} coherent negative pressure-strain interaction features that indicate the growth and overshoot phases of reconnection, {\it i.e.,} the interval during which the reconnection rate is rapidly increasing following the slow onset phase and followed by the time interval before reaching a steady-state.} 
We present a physical marker indicating the early phases of reconnection during which the reconnection rate is rapidly increasing following the slow onset phase.  The marker is a coherent strong positive signal of pressure-strain interaction sandwiched between negative signals.
%\textcolor{red}{We present a coherent strong positive signal of pressure-strain interaction sandwiched between negative signals as a physical marker indicating the early phases of reconnection during which the reconnection rate is rapidly increasing following the slow onset phase.}
The key physics is that the separatrix opening angle increases relatively rapidly inside the EDR as the electrons deflect around the pre-existing downstream current sheet and/or the magnetic island generated during the onset phase. This diverging flow leads to a negative feature in the electron pressure-strain interaction that appears near the downstream edge of the EDR, displaced from the neutral line by approximately one electron inertial scale. 
These negative features appear alongside a strong positive signal of pressure-strain interaction resulting from the compression that occurs at a dipolarization front.
%\textcolor{red}{As flows get deflected around, these negative features appear with a strong positive signal near the neutral line which is a marker for dipolarization fronts.}
%The $-{\bf P}_e:{\bf S}_e<0$ feature has an apparent motion in the downstream direction once a steady-state has been reached in the EDR.  
%The negative features when accompanied by a strong positive signal sandwiched between them serve as a
This pattern in $-{\bf P}_e:{\bf S}_e$ is a good marker for the reconnection growth and overshoot phases because they only appear in the EDR's surroundings during these phases and are absent during the onset and steady-state phases. The quantity lends itself to measurement by spacecraft because it is a scalar that does not require identification of the reconnection geometry, and uses well-established quantities measurable by MMS \cite{Chasapis18,Bandyopadhyay20,Burch_PoP_2023}.

The underlying mechanism was initially hypothesized in a previous study \cite{Barbhuiya_PiD3_2022}, but could not be tested numerically because the simulation system size was too small to achieve a steady-state phase. In this study, we validate the theory using a larger simulation domain that reaches a steady-state. We demonstrate that the location, extent, amplitude, and time evolution of the marker align with expectations. 

It is crucial to emphasize that the electron pressure-strain interaction feature described here is insufficient in isolation to imply that reconnection is happening.
%and is in its \textcolor{red}{growth/overshoot phases}.  
Rather, this signature must be used in concert with other standard identifiers of reconnection, including jets and reversing magnetic fields. Moreover, as the crossing events by spacecraft should be close to the EDR, quantities that are necessary to establish if an EDR is observed, \textit{e.g.,} agyrotropy \cite{Swisdak_2016_GRL}, should also be employed in tandem.

\begin{figure}
    \begin{center}
    \includegraphics[width=3.2in]{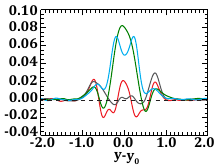}
    \caption{1D cuts of the electron pressure-strain interaction along the four vertical dashed lines at $t=18$ [panel (c) of Fig.~\ref{fig:2D_PS_NegEllipses}] to simulate potential spacecraft observations. The cuts are color-coded to match the respective dashed lines. This highlights the variability of the $-{\bf P}_e:{\bf S}_e$ signal.} 
    \label{fig:1Dcut_t18} 
    \end{center}
\end{figure}

A second key point is that both the positive and negative features in pressure-strain interaction may or may not be captured
%It is conceivable that not all the characteristics of the described electron pressure-strain interaction marker, \textit{i.e.,} a positive signal flanked by negative signals, can be captured 
in a single spacecraft flyby through single-point spatial measurements or the barycenter of a four spacecraft constellation such as MMS. 
%Thus, we emphasize that
%%We posit that 
%multiple single-point measurements may be a necessity for a clear identification of the proposed marker.
%%Therefore, to more 
%To demonstrate this using simulation data, we plot the traces of virtual satellites to
%%effectively 
%simulate a crossing event by MMS.
%%using multiple satellites, we use the example of}
%\textcolor{orange}{We simulate a crossing event for MMS.
%%(a constellation of four satellites). 
We plot four virtual spacecraft trajectories separated by approximately $1 \ d_e$ in perfectly vertical cuts normal to the horizontal neutral line through $-{\bf P}_e:{\bf S}_e$ at the peak growth time $t=18$, demonstrated in Fig.~\ref{fig:2D_PS_NegEllipses}(c). The gray, red, green, and blue lines are at $x-x_0 = -3.8, -3.3, -2.8,$ and $-2.3$. Fig.~\ref{fig:1Dcut_t18} displays 1D cuts along these four lines, highlighting the different 1D signatures along the virtual spacecraft trajectories.
%, \textcolor{red}{and thereby simulating possible different single point measurements.} 
The blue line at -2.3 captures only the strong positive $-{\bf P}_e:{\bf S}_e$ signal due to the dipolarization front, with no negative $-{\bf P}_e:{\bf S}_e$ feature, while the green line at -2.8 captures the positive signal and part of the coherent negative $-{\bf P}_e:{\bf S}_e$ feature. The red line at -3.3 displays the dominant negative $-{\bf P}_e:{\bf S}_e$ feature yet not capturing much of the $-{\bf P}_e:{\bf S}_e>0$ signal.
The gray line at -3.8 barely detects any of the coherent $-{\bf P}_e:{\bf S}_e$ signal.
This example demonstrates that a single trajectory may or may not detect both the positive and negative features in the electron pressure-strain interaction.

For the negative electron pressure-strain interaction feature to be a useful identifier of the growth or overshoot phases of reconnection in nature, it needs to be resolvable using measurement devices.  To assess whether the feature would be measurable using the MMS satellites, we consider the well-studied 11 July 2017 EDR crossing event in the magnetotail \cite{Torbert18} since magnetotail reconnection is reasonably symmetric and anti-parallel, as in the simulations used in the present study. From Fig.~\ref{fig:1Dcutsallstages}, the localized negative $-{\bf P}_e:{\bf S}_e$ features have an amplitude of $\approx 0.01-0.05 \ (B_0^2/4 \pi)\Omega_{ci0}$ in our simulations. 
The simulation employed in this study uses an unrealistic ion-to-electron mass ratio of 25, lower than the realistic value by a factor of 73. In Ref.~\cite{Barbhuiya_PiD3_2022}, it was argued in their Eq.~(8) that the electron pressure-strain interaction scales as  inverse of the electron mass. Then, the pressure-strain feature for a realistic electron mass would be $\approx 0.73-3.65 \ (B_0^2/4 \pi)\Omega_{ci0}$.  Using $B_0 \simeq 12$ nT \cite{Argall22}, we obtain $(B_0^2/4 \pi)\Omega_{ci0} = 0.13 \ {\rm nW/m^3}$. Finally, we obtain that the negative $-{\bf P}_e:{\bf S}_e$ features would have an amplitude of $\approx 0.095-0.475~{\rm nW/m^3}$ for the 11 July 2017 EDR crossing.
Signals of pressure-strain interaction at the level of $0.2~{\rm nW/m^3}$ were previously detected by MMS for this event, as reported in recent work \cite{Burch_PoP_2023}, so we predict that the negative $-{\bf P}_e:{\bf S}_e$ features discussed here, which are weaker than the strong positive feature, are in principle measurable by MMS.

This study focused on the pressure-strain interaction of electrons. For ions, a similar signature would be expected at the edge of the IDR. However, the strength of such a signature scales as $P_{i,up} \Omega_{ci,up}$ \cite{Barbhuiya_PiD3_2022} instead of $P_{e,up} \Omega_{ce,up}$ for the electrons, where 
%its electron counterpart. 
%Here, 
$up$ means upstream of the IDR for ions ($i$) and upstream of the EDR for electrons ($e$). Assuming a magnetotail-relevant $P_{i,up} \simeq 5P_{e,up}$, we find that the ion signal would be a factor of 50 smaller than the electron signal, or of the order of %$0.002~{\rm nW/m^3}$.
$0.01~{\rm nW/m^3}$. This is unlikely to be measurable with certainty by MMS.

We note that the physical marker discussed here should be relatively robust, as it should occur with or without an out-of-plane (guide) magnetic field and for both symmetric or asymmetric reconnection, because the separatrices in the downstream region open out during the growth phase. However, the spatial structure of the negative pressure-strain interaction is likely different, as a (strong) guide field introduces a quadrupolar pattern in the electron pressure \cite{Kleva95} and an asymmetry leads to the opening of the separatrix preferentially to the low magnetic field side [{\it e.g.,} \cite{cassak07a}].  In the presence of an upstream flow shear, such as reconnection at Earth's dayside magnetopause \cite{gosling86a,Gosling91}, the feature in pressure-strain interaction discussed here may be overcome by the pressure-strain interaction associated with the upstream flow shear, which could render this marker of the growth phase less important in this setting. For current sheets with secondary islands, the signature may not work as cleanly, as was demonstrated by Fig.~\ref{fig:2D_PS_NegEllipses}(c). Reconnection is also a significant element of turbulence [\textit{e.g.}, \cite{Servidio_PRL_2009}], and turbulence often involves flow shear between interacting eddies. The presence of flow shear could similarly complicate the use of the marker discussed here at reconnection sites embedded in a turbulent system. 

Though a robust indicator, this marker has certain limitations. The overshoot phase discussed in this manuscript may or may not 
%necessarily 
occur in Nature as it does in the simulation used here. If absent in Nature, the presence of the pressure-strain interaction marker could indicate not only the growth phase but also the early portion of the
%what could be regarded as a 
steady-state phase. However, as shown in panel (g) of Fig.~\ref{fig:2D_PS_NegEllipses}, if this marker were present during the steady-state phase, it would be located well outside the EDR. Future work should employ a
%involve 
simulation without an overshoot phase to check for the presence of pressure-strain markers during the growth and steady-state phases.

As this study uses a single magnetotail-relevant simulation, future work should vary upstream parameters like species' temperatures, temperature ratio, number density, guide field strength, and mass ratio to assess how these affect the feature's strength and structure. Including asymmetries and upstream flow shear will be valuable for dayside magnetopause and solar wind conditions. Examining 3D effects is also crucial. For instance, in guide field reconnection with a diffusion region that is localized in the out-of-plane direction, the separatrices open out in the inflow direction in the diffusion region but open in the out-of-plane direction downstream of it, \cite{sasunov2015statistical,shepherd2017structure}, which undoubtedly impacts the structure and dynamics of the pressure-strain interaction features discussed in the present study. Exploring the marker's usefulness in turbulent systems, relevant to the magnetosheath and solar wind, is also important. Ultimately, it is essential to investigate whether this marker has been observed by spacecraft like MMS.

\acknowledgments
We acknowledge helpful conversations with Matthew Argall. PAC, AC, and MAS gratefully acknowledge primary support from NASA grant 80NSSC24K0172. PAC acknowledges supplemental support from NSF grant PHY-1804428, DOE grant DE-SC0020294, and NASA grant 80NSSC19M0146  PAC also gratefully acknowledges support and hospitality during his visit as an Invited Senior Researcher at the Laboratoire de Physique des Plasmas École Polytechnique.  MAS also acknowledges support from NSF grant AGS-2024198 and NASA grant 80NSSC20K1813. GC is supported by the Academy of Finland Grant n.345701. This research was partially supported by the International Space Science Institute (ISSI) in Bern, through ISSI International Team project \#23-588 (``Unveiling Energy Conversion and Dissipation in Non-Equilibrium Space Plasmas'').  This research used resources of the National Energy Research Scientific Computing Center (NERSC), a U.S. Department of Energy Office of Science User Facility located at Lawrence Berkeley National Laboratory, operated under Contract No. DE-AC02-05CH11231 using NERSC award FES-ERCAP0027083.  Simulation data used in this manuscript are available on Zenodo (https://doi.org/10.5281/zenodo.13858655).

\bibliography{BlueEllipses}

\end{document}